\documentclass[aps,preprint,showpacs,preprintnumbers,amsmath,amssymb,floatfix]{revtex4-1}

\usepackage{graphicx}
\usepackage{dcolumn}
\usepackage{bm}
\usepackage{CJK}

\begin{document}

\title{Violation of the virial theorem and generalized equipartition theorem for logarithmic oscillators serving as a thermostat}

\author{Kai Chen}
\affiliation{Department of Physics and Institute of Theoretical Physics and Astrophysics, Xiamen University, Xiamen 361005, China}
\author{Dahai He}\email[]{dhe@xmu.edu.cn}
\affiliation{Department of Physics and Institute of Theoretical Physics and Astrophysics, Xiamen University, Xiamen 361005, China}
\author{Hong Zhao}
\affiliation{Department of Physics and Institute of Theoretical Physics and Astrophysics, Xiamen University, Xiamen 361005, China}
\affiliation{Collaborative Innovation Center of Chemistry for Energy Materials, Xiamen University, Xiamen 361005, China}

\date{\today}
\begin{abstract}
A logarithmic oscillator has been proposed to serve as a thermostat recently since it has a peculiar property of infinite heat capacity according to the virial theorem.  In order to examine its feasibility in numerical simulations, a modified logarithmic potential has been applied in previous studies to eliminate the singularity at origin. The role played by the modification has been elucidated in the present study. We argue that the virial theorem is practically violated in finite-time simulations of the modified log-oscillator illustrated by a linear dependence of kinetic temperature on energy. Furthermore, as far as a thermalized log-oscillator is concerned, our calculation based on the canonical ensemble average shows that the generalized equipartition theorem is broken if the temperature is higher than a critical temperature. Finally, we show that log-oscillators fail to serve as thermostats for its incapability of maintaining a nonequilibrium steady state even though their energy is appropriately assigned.
\end{abstract}
\maketitle
\thispagestyle{empty}
\section*{Introduction}\label{sec:1}
Theoretical and numerical investigations of statistical mechanical systems invariably rely upon a suitable modeling of thermostats. Two commonly used models in literature are Langevin thermostat and Nos\'{e}-Hoover thermostat (see, e.g., \cite{coffey2012, Hoover2012book} and references therein). The former involves in cumbersome calculations of stochastic processes and the latter is a deterministic bath with time-reversible dynamics, which however can give rise to irreversible dissipative behavior. A new model of thermostat, capable of producing controlled dynamics to sample canonical ensemble, is always desirable from both theoretical and numerical viewpoints.

Recently, a logarithmic oscillator (log-oscillator) has been suggested to be a candidate to serve as a Hamiltonian thermostat~\cite{campisi2012PRL}. In the simplest 1D version, the Hamiltonian of a log-oscillator reads
\begin{equation}\label{Hamiltonian1}
H_{log}=\frac{p^{2}}{2m}+k_B\tau\ln\frac{|x|}{b},
\end{equation}
where $p$, $x$, $m$ are the momenta, position and mass of the log-oscillator, respectively. The positive constant $b$ sets the length scale of the oscillator. $\tau$ is the \textit{thermostat temperature} and $k_B$ the Boltzmann's constant. According to the virial theorem~(see \cite{tuckerman2010statistical})  for the log-oscillator
\begin{equation}\label{virial-origin}
\left \langle p\frac{\partial H_{log}}{\partial p} \right \rangle= \left \langle x\frac{\partial H_{log}}{\partial x}\right \rangle,
\end{equation}
where $\langle\cdot\rangle$ denotes the time average taken with respect to a dynamical trajectory, one immediately has $\langle\frac{p^{2}}{m}\rangle=k_{B}\tau$, which means the kinetic temperature $T_{k}\equiv\frac{1}{k_{B}}\langle\frac{p^{2}}{m}\rangle$ is always equal to $\tau$ for all trajectories regardless of their energy $E$. Therefore, $\frac{\partial T_{k}}{\partial E}=0 $, indicating an infinite heat capacity~\cite{campisi2012PRL, campisi2013JPCB}. This implies that its kinetic temperature can maintain a constant value even though the log-oscillator couples with another system, which absorbs energy from the log-oscillator or vice versa. This peculiar property make the log-oscillator an ideal candidate of thermostats. Although the log-oscillator has the outstanding property, its feasibility as a thermostat has been argued recently~\cite{melendez2012comment,campisi2012reply,hoover2013CNSNS,sponseller2014PRE}. The arguments mainly lie in its long length and time scales involved in practical implementations or numerical simulations.  It has particularly been shown that the log-oscillator is incapable of establishing the heat flow in a nonequilibrium system~\cite{hoover2013CNSNS} and fails to serve as a thermostat for small atomic clusters~\cite{sponseller2014PRE}.

Actually, it has been known for a long time that Brownian particles with momentum-dependent drift can be described by power-law tail distributions and correspondingly show anomalous diffusion in optical lattices~\cite{Lutz2004PRL,Barkai2010PRL,Ori2011PRE,Baikai2012PRE,Barkai2014PRX,Barkai2015PRLa,Lutz2015PRL,Barkai2015PRLb,Barkai2016PRE}. In this case, systems are equivalently described by Brownian particles confined inside a logarithmic potential in momentum space. Ergodicity of such systems is accordingly broken~\cite{Lutz2004PRL,Baikai2012PRE} and the Brownian motion approaches an infinite covariant density~\cite{Barkai2010PRL}. Distinct phases of the anomous diffusion and nonstationary scale-invariant dynamics have been extensively investigated~\cite{Ori2011PRE,Baikai2012PRE, Barkai2014PRX,Barkai2015PRLa,Lutz2015PRL}. The phase-space density can deviate from the Boltzmann-Gibbs statistics at the intermediate energy and exhibit heavy power-law tails~\cite{Barkai2015PRLb,Barkai2016PRE}.

Note that in order to avoid the singularity at the origin, a common way in previous studies is to apply a modified form of the logarithmic potential (see Eq.~\eqref{Hamiltonian2} below) when one performs numerical simulations~\cite{campisi2012PRL,hoover2013CNSNS,campisi2013JPCB,sponseller2014PRE}. Some controversial conclusions on the original log-oscillator~\eqref{Hamiltonian1} given in previous studies were drawn from the simulation results based on the modified log-oscillator. However, less attention has been paid to the effects coming from the modification required in numerical studies. It is therefore necessary to elucidate the role played by the modification of the logarithmic potential and further understand the underlying mechanism for the failure of modified log-thermostat in numerical simulations. In the present study, we study the statistical property of the modified log-oscillator from both microcanonical and canonical aspects. Interestingly, we find that the virial theorem is seemingly violated, illustrated by an unexpected energy dependence of kinetic temperature numerically. Furthermore, as a thermalized log-oscillator is concerned, our analysis shows the generalized equipartition theorem is broken when the temperature is higher than a critical temperature, which is verified by numerical simulations. Finally, we show that modified log-oscillators cannot serve as thermostats to produce a stationary temperature profile inside the FPU-$\beta$ system even though the energy of log-oscillators is appropriately assigned.
\section*{Results and Discussion}\label{sec:2}
\subsection*{Isolated log-oscillator: Violation of the virial theorem}
For the sake of avoiding the singularity of the logarithmic potential at $x=0$ particularly when one performs numerical simulations, as mentioned above, an usual way is replacing the logrithmic potential of Eq.~\eqref{Hamiltonian1} with the following modified potential~\cite{campisi2012PRL,hoover2013CNSNS,campisi2013JPCB,sponseller2014PRE}:
\begin{equation}\label{Hamiltonian2}
V_{log}=\frac{k_B\tau}{2}\ln\left[\left(\frac{x}{b}\right)^{2}+\varepsilon\right].
\end{equation}
One can see that the presence of the parameter $\varepsilon$ eliminates the singularity and make numerical simulations feasible. However, some peculiar effects come from the presence of $\varepsilon$ as shown below. The equation of motion is solved by using the velocity-Verlet algorithm and a time step of 0.0001. This algorithm is symplectic and energy conservation is numerically satisfied in our simulations. In what follows, $m$, $k_B$ and $b$ are set equal to unity unless otherwise stated. The results of numerical simulation shown in this section implying that the small parameter $\varepsilon$ lead to violation of the virial theorem for the log-oscillator.

By defining \textit{kinetic temperature} $T_k\equiv\frac{1}{k_{B}}\langle\frac{p^2}{m}\rangle$, the virial theorem for the modified system~\eqref{Hamiltonian2} yields
\begin{equation}\label{virial-modified}
T_{k}=\tau \left\langle \frac{x^2}{x^{2}+\varepsilon} \right\rangle\equiv P_{k},
\end{equation}
where $\langle\cdot\rangle$ refers to a long-time average with respect to a dynamical trajectory.
\begin{figure}[htbp]
\centering
\includegraphics[width=0.75\columnwidth]{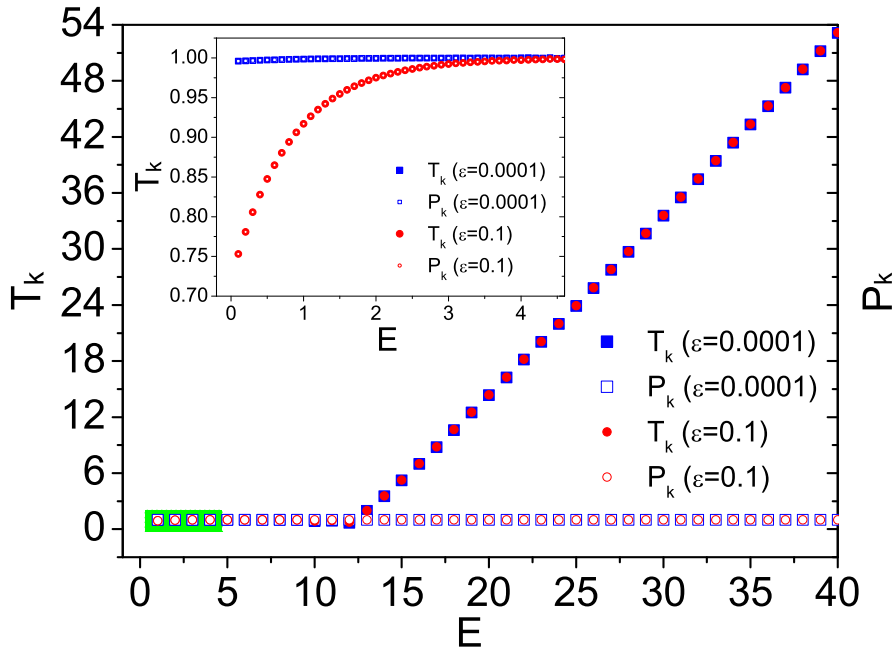}
\caption{\label{Fig1} Kinetic temperature $T_{k}$ and the quantity $P_{k}$ as a function of the total energy $E$ of the isolated log-oscillator for different $\varepsilon$. In our simulation, $\tau=1$, relaxation time and average time are $10^{8}$ and $2\times10^{9}$ time steps, respectively. Numerical fitting gives $T_{k}=2(E-E_{c})$ for the linearly increasing part. Inset: Enlarged view of low-energy part within the green rectangle.}
\end{figure}
\begin{figure}[htbp]
\centering
\includegraphics[width=0.7\columnwidth]{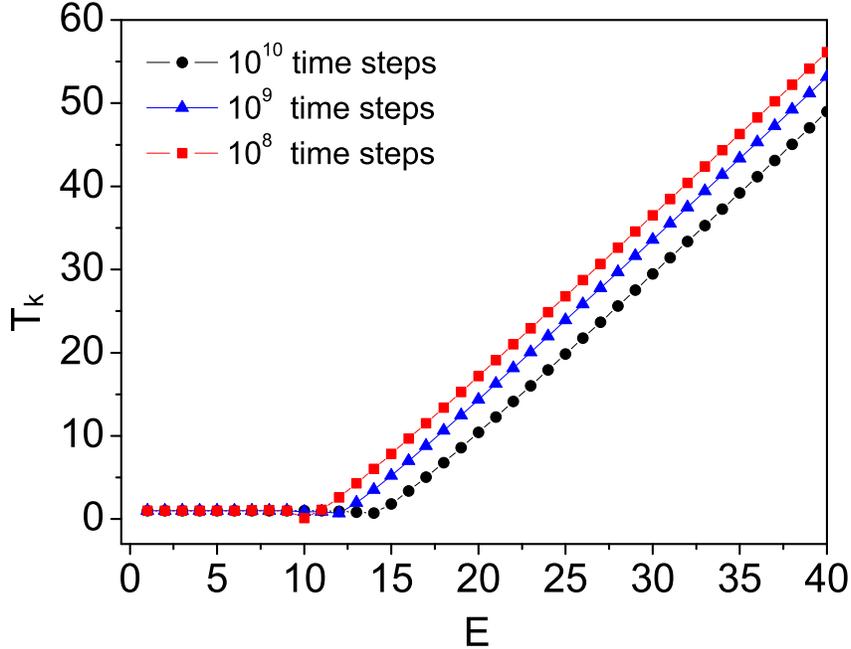}
\caption{\label{Fig2} Kinetic temperature $T_{k}$ as a function of the total energy $E$ of the isolated log-oscillator for different average time. Here $\tau = 1$ and $\varepsilon = 0.0001$. The relaxation time is $10^{8}$ time steps for all three cases.}
\end{figure}

For an ideal log-oscillator, kinetic temperature $T_{k}$ should be energy independent according to the virial theorem. As a small $\varepsilon$ is introduced in a way like Eq.~\eqref{Hamiltonian2}, energy dependence of $T_{k}$ is notable only when energy is very low and comes to vanish as energy increases. Surprisingly, we observe an anomalous energy dependence of $T_{k}$ which seemingly deviates from the prediction from virial theorem when energy of the system is large. As shown in Fig.~\ref{Fig1}, we plot the dependence of kinetic temperature $T_{k}$ on the total energy $E$ of the system. One can see that there is a plateau region for a certain range of energy, indicating that the kinetic temperature of log-oscillator is equal to  $\tau$ and independent on the energy of system. In this region, the log-oscillator nearly has an infinite heat capacity. However, $T_k$ increases linearly with $E$ as $E>E_c$, where the critical energy $E_c$ is independent of the magnitude of $\varepsilon$.  Our numerical studies show that the plots coincide for different value of $\varepsilon$, even for the case that $\varepsilon$ is as small as $10^{-8}$. For the increasing part, linear fitting gives that $T_k=2(E-E_c)$, which indicates $E_{c}=\langle V_{log}\rangle$. The fitting relation $T_k=2(E-E_c)$ simply comes from the energy conservation of the Hamiltonian system. Unlike a ``normal" thermodynamical system, for which the potential energy quickly obtains the increment of energy while the kinetic energy is kept constant in average, the log-oscillator has a slow relaxation of energy. This explains why the kinetic temperature is proportional to the energy since the potential energy in average gains little the energy increment while kinetic energy gains almost all of them. Note that when energy $E$ is small, as shown in the inset of Fig.~\ref{Fig1}, $T_k$ deviates from the value of $\tau$ while the virial theorem, i.e., $T_{k}=P_{k}$, is valid in this case as expected. The deviation is more considerable as $\varepsilon$ increases. The presence of $\varepsilon$ leads to the correction in the density of states in the low energy regime (see, e.g., Fig.~2 in Ref.~\cite{campisi2012PRL}). To reduce the deviation, it has been argued that the total energy $E$ should be large~\cite{campisi2012PRL, melendez2012comment}. However, our results indicates that total energy should not be ``too large", i.e., $E$ should be less than $E_{c}$. Otherwise, the modification of logarithmic potential seemingly indicates a violation of virial theorem, which is required for the log-oscillator to serve as a thermostat.

We also study the effect of average time on the behavior of $T_k$, as shown in Fig.~\ref{Fig2}. One can see that the plateau region is broaden with the increase of average time. However, the broaden process is extremely slow. Note that the critical energy $E_c$ shifts roughly a constant amount through increasing the simulation time by a factor of 10, which indicates an exponential dependence of the required simulation time on energy. As the integration goes to infinity, it is unrealistic through numerics to conclude that whether the broaden process would stop at some critical energy (similar as the \textit{critical temperature} given in the section below) or keep ongoing, which requires a solid analytical study. However, since the model described by Eq.~\eqref{Hamiltonian2} is proposed mainly for serving as a thermostat for numerical simulations, it is safely to argue that virial theorem is practically ``violated" for the modified log-oscillator as $E>E_c$, which results in its failure to serve as a thermostat.

\begin{figure}[htbp]
\centering
\includegraphics[width=0.85\columnwidth]{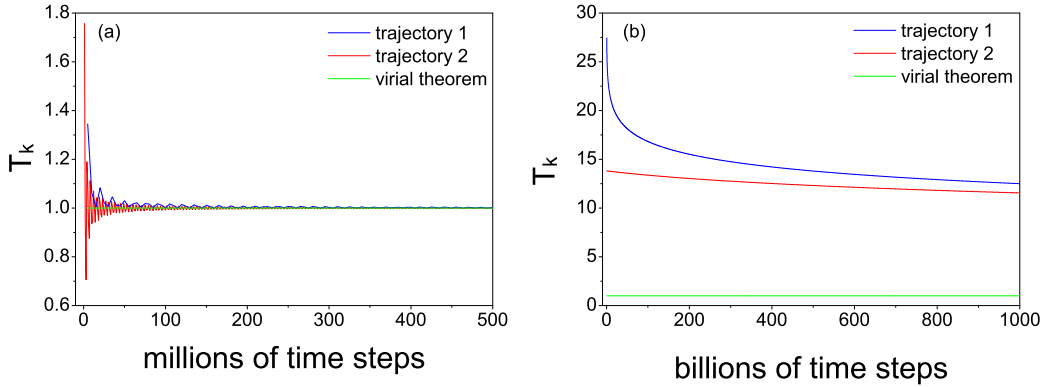}
\caption{\label{Fig3} $T_{k}$ as a function of average time for an isolated oscillator with the total energy (a) $E=5$ and (b) $E=25$. The average is along two typical trajectories (see text for details) denoted by the blue line and the red line, respectively.  The prediction by the virial theorem ($T_{k}=P_{k}$) is drawn by green lines as a reference. Here $\tau=1$, $\varepsilon=0.0001$, the relaxation time is given by $10^{8}$ time steps. Note that the time unit of the right plot is billions ($10^9$) of time steps. For such a long-time simulations, the energy drift in our simulations is controlled in the order of $10^{-4}$ (trajectory 1) and $10^{-7}$ (trajectory 2).}
\end{figure}
In order to understand the underlying mechanism further, we study the relaxation process of $T_{k}$ by setting the log-oscillator's total energy inside the plateau region ($E=5$) and the linear increase region ($E=25$), respectively. Meanwhile, we choose two typical trajectories for which the log-oscillator is located either near the origin (trajectory 1) or far from origin (trajectory 2) initially. For the case $E=5$, $T_{k}$ relaxes quickly to the predicted value in terms of the virial theorem (Fig.~\ref{Fig3}(a)). However, for the case $E=25$, the relaxation process is extremely slow shown by our long-time simulations up to $10^{12}$ time steps (Fig.~\ref{Fig3}(b)). Its relaxation to the equilibrium state is unavailable through practical simulations. The slow relaxation lies in the exponential growing of oscillation time with the energy of the log-oscillator. Therefore, in a much shorter timescale attainable in numerical simulations, the log-oscillator essentially moves with a constant velocity, which explains the linear behavior with energy shown in Fig.~\ref{Fig1} and Fig.~\ref{Fig2}.

As to make sure the violation does not originate from numerical errors, we also check our results by two higher-order symplectic algorithms, namely, Omelyan-Mryglod-Folk algorithm~\cite{Omelyan2002CPC} and Laskar-Robutel algorithm~\cite{Lascar2001CMDA}. The two algorithms have four-order and ten-order symplectic integrators, respectively. They recover the results by the velocity-Verlet algorithm with the total-energy drift as small as $10^{-10}$ in a trillion-timestep run, namely, the accuracy is increased by three to four orders of magnitude in comparison with the Verlet algorithm.
\subsection*{Thermalized log-oscillator: Violation of the generalized equipartition theorem}\label{sec:3}
Let us revisit the so-called generalized equipartition theorem (GET) originally derived by Tolman~\cite{Tolman1918GET}, which is represented by the relation
\begin{equation}\label{GET}
\langle x_{i} \frac{\partial H}{\partial x_{j}}\rangle = \delta_{ij}k_{B}T. \\
\end{equation}
Here $H$ is the Hamiltonian of a system, $x_{i,j}$ stand for canonically conjugate (generalized) coordinates or (generalized) momentums. In this section, $\langle\cdot\rangle$ denotes the canonical ensemble average at equilibrium temperature $T$. A special case of GET is the well-known equipartition theorem for a system of quadratic terms only. GET ranks among the basic results of statistical mechanics and has been extensively studied until today~\cite{Uline2008JCP, Claudio2014GET}. Note that although GET is similar with the virial theorem (see, e.g., Eq.~\eqref{virial-origin} or Eq.~\eqref{virial-modified}) in form, canonical ensemble average is applied here instead of time average.

For a one-dimensional system of Hamiltonian $H= \frac{p^{2}}{2m} + V(x)$, Eq.~\eqref{GET} indicates $\langle x \frac{\partial V}{\partial x}\rangle = \langle \frac{p^{2}}{m}\rangle = k_{B}T$. In thermal equilibrium state, the canonical distribution function is given by $\rho(x, p) = \frac{1}{Z}e^{-\beta H(x, p)} = \frac{1}{Z_{p}}e^{-\beta \frac{p^{2}}{2m}} \frac{1}{Z_{x}}e^{-\beta V(x)}$, where $\beta = 1/k_{B}T$, $Z_{p}$ and $Z_{x}$ are reduced participation functions with respect to $p$ and $x$, respectively. In the section above we show that the virial theorem is practically violated for the log-oscillator when we replace the logarithmic potential by Eq.~\eqref{Hamiltonian2}. In this section we will analytically show that GET is also broken for the modified log-oscillator and verify the result by numerical simulations.

On one hand, according to Eq.~\eqref{GET} with respect to the momentum coordinate $p$, one immediately obtains
\begin{equation}\label{Ke}
K_{\mu} \equiv <p \frac{\partial H}{\partial p}> = \frac{1}{Z_{p}}\int_{-\infty}^{+\infty} dp \frac{p^{2}}{m} e^{-\beta \frac{p^{2}}{2m}} =k_{B}T. \\
\end{equation}
By defining a canonical temperature~\cite{Bachmann2014Tem} $k_{B}T_{\mu}\equiv K_{\mu}$, one have $T_{\mu} = T$, showing the validity of GET with respect to $p$.

On the other hand, Eq.~\eqref{GET} with respect to position coordinate $x$ yields
\begin{equation}\label{Pe}
P_{\mu} \equiv <x \frac{\partial H}{\partial x}> = \frac{1}{Z_{x}}\int_{-\infty}^{+\infty} dx x \frac{\partial H}{\partial x} e^{-\beta V(x)}.\\
\end{equation}
Calculations shows that there exists a critical temperature $T=\tau$ for the integration above, for which one have $P_{\mu}=k_{B} T$ is valid only for $T<\tau$ (see details in the section of methods). $P_{\mu}$ deviates from the linear behavior and approaches to an upper bound $k_B\tau$ for $T\ge\tau$. This means that GET with respect to coordinate $x$ is broken.
\begin{figure}[htbp]
\centering
\includegraphics[width=1\columnwidth]{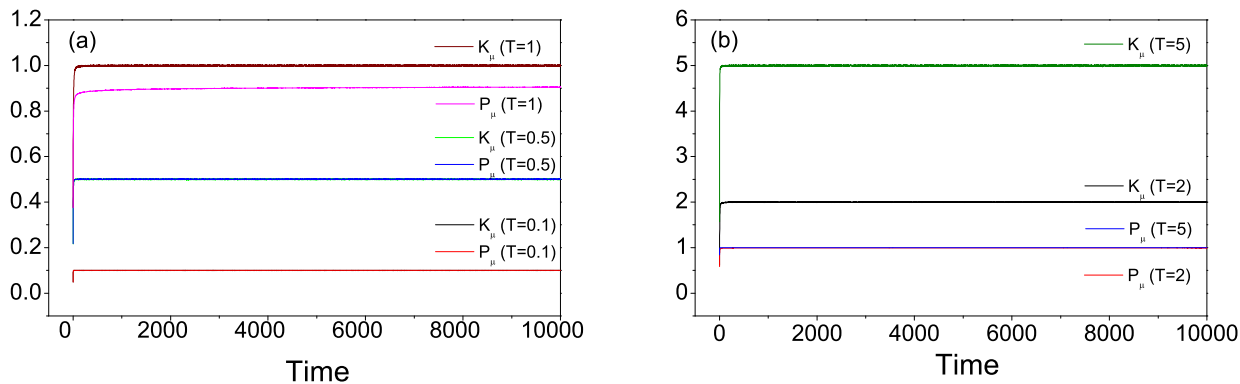}
\caption{\label{Fig4} $K_{\mu}$ and $P_{\mu}$ as function of evolution time for different Langevin thermostat's temperature $T$. For a given time, $K_{\mu}$ and $P_{\mu}$ are calculated by ensemble average. Here $\tau = 1$, $\epsilon=0.0001$, $\gamma=0.7$, and the number of ensembles is $1\times10^{6}$.}
\end{figure}

In order to verify the above analytical results, we perform numerical simulations by thermalizing the log-oscillator with a Langevin thermostat at temperature $T$. The equation of motion is given by
\begin{eqnarray}\label{eom-langevin}
m\ddot{x}=-\frac{\partial V_{log}}{\partial x}-\gamma\dot{x}+\eta,
\end{eqnarray}
where $\gamma$ is the friction constant and $\eta$ denotes a Gaussian white noise with zero mean and variance of $2\gamma k_{B}T$. Eq.~\eqref{eom-langevin} is integrated by using a stochastic velocity-Verlet algorithm.
\begin{figure}[htbp]
\centering
\includegraphics[width=0.7\columnwidth]{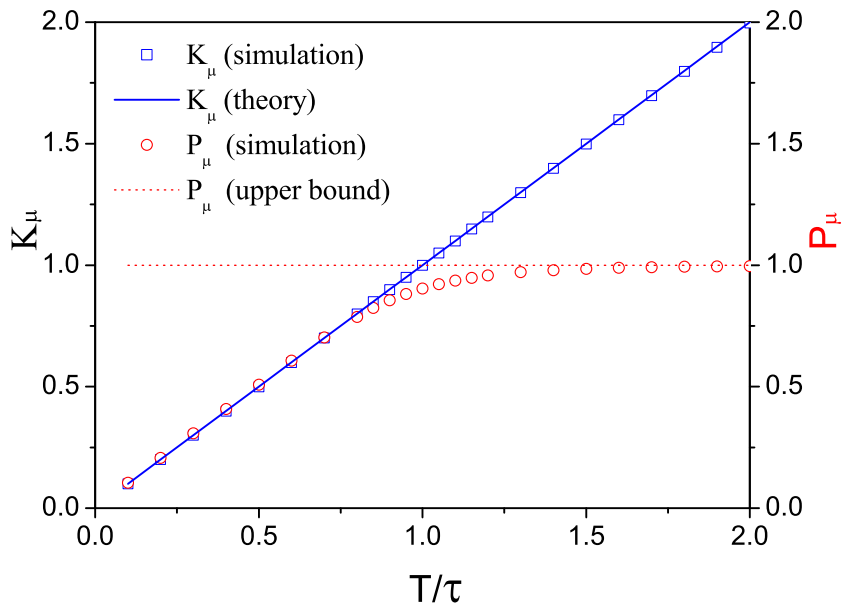}
\caption{\label{Fig5} $K_{\mu}$, $P_{\mu}$ as function of the scaled temperature $T/\tau$. Here $10^6$ ensembles are used and each ensemble of the system evolves $10^6$ time steps.  Here we set $\tau = 1$, $\epsilon =0.0001$, and $\gamma=0.7$.}
\end{figure}

Firstly, we study the dynamical behavior of $K_{\mu}$ and $P_{\mu}$ for some typical equilibrium temperatures, as shown in Fig.~\ref{Fig4}, where a canonical ensemble average is taken for any given time. For $T=0.1$, after a quick relaxation process, $K_{\mu}$ and $P_{\mu}$ saturates at the value predicted by GET, i.e., $K_{\mu}=P_{\mu}=k_{B}T$. This is also the case for $T=0.5$ with a slightly slower relaxation process. For $T\geq\tau$, while $K_{\mu}$ keep satisfying GET, one cannot see a sign for $P_{\mu}$ converging to GET prediction. One can see the transition clearly from Fig.~\ref{Fig5}, where we plot $K_{\mu}$ and $P_{\mu}$ as a function of thermostat temperature. Unlike $K_{\mu}$, $P_{\mu}$ approaches to the upper bound $k_B\tau$ as $T$ increases, which is consistent with the analysis above. Note that the transition of $P_{\mu}$ near $T=\tau$ is not sharp but smoothly curvy. The deviation from the theoretical prediction may come from finite number of canonical sampling and finite relaxation time in our simulations. Our simulations show that $P_{\mu}$ will gradually close to the analytical value by increasing the relaxation time.
\subsection*{Failure of log-thermostat to maintain a nonequilibrium steady state}\label{sec:4}
In this section, we apply the log-thermostat to see if it can drive the system into a nonequilibrium state. It has been shown that Hamiltonian thermostats, including the log-thermostat, fail to promote heat flow~\cite{hoover2013CNSNS}. Although there is a linear energy dependence for $T_{k}$ when $E>E_{c}$, it would be still interesting to learn if the log-oscillator can serve as a thermostat when its energy is \textit{appropriately} set inside the plateau region (Fig.~\ref{Fig1}). It seems promising since $T_{k}$ is practically energy independent and the log-oscillator nearly has infinite heat capacity in this case.

As the usual setup for heat conduction, the system is sandwiched between two log-thermostats at different temperatures through a weak coupling. The whole Hamiltonian is given by
\begin{equation}
   H=H_{s}+H^{\pm}_{log}+H^{\pm}_{int}.
\end{equation}
The system of interest is a one-dimensional FPU-$\beta$ chain of oscillators
\begin{equation}
   H_{s}=\sum_{n=2}^{N-2}\frac{p_{n}^{2}}{2}+V(x_{n+1}-x_{n}), \\
\end{equation}
where $V(y)=\frac{1}{2}y^{2}+\frac{1}{4}y^{4}$. The bath Hamiltonian $H^{\pm}_{log}$ are
 \begin{equation}
   \begin{aligned}
   H^{-}_{log}&=\frac{p_{1}^{2}}{2}+\frac{\tau_{-}}{2}\ln(x_{1}^{2}+\varepsilon),\\
   H^{+}_{log}&=\frac{p_{N}^{2}}{2}+\frac{\tau_{+}}{2}\ln(x_{N}^{2}+\varepsilon).\\
   \end{aligned}
\end{equation}
The system-bath interaction Hamiltonian $H^{\pm}_{int}$ are given in a quadratic form~\cite{sponseller2014PRE}
\begin{equation}
  \begin{aligned}
   H^{-}_{int}&=\frac{\gamma_{-}}{2}(x_{2}-x_{1})^{2},\\
   H^{+}_{int}&=\frac{\gamma_{+}}{2}(x_{N}-x_{N-1})^{2}, \\
   \end{aligned}
   \end{equation}
where $\gamma_{\pm}$ are the strengths of the couplings between the log-thermostats and the FPU-$\beta$ system.

\begin{figure}[htbp]
\centering
\includegraphics[width=1\columnwidth]{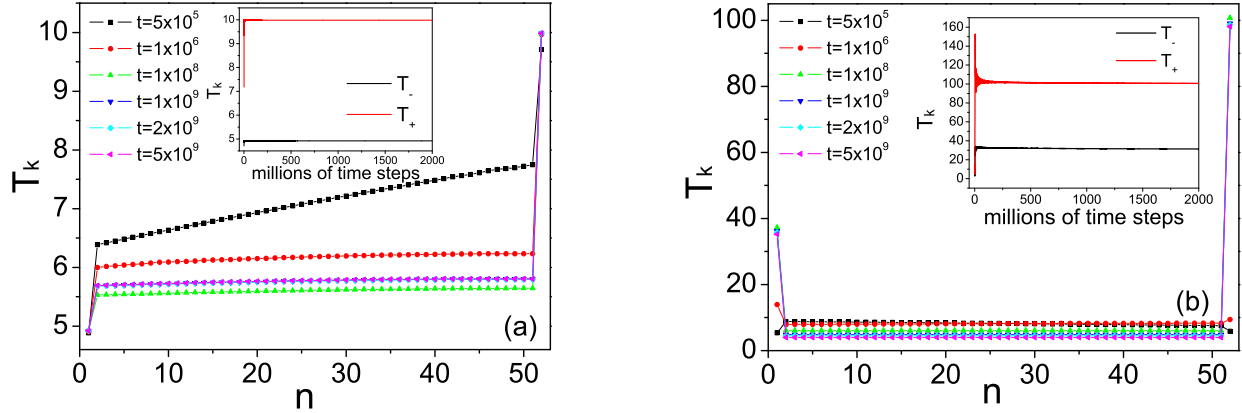}
\caption{\label{Fig6} Temperature profiles different average time. Energy of respective log-oscillators is initially given by (a) $E_{-}=1$ and $E_{+}=20$; (b) $E_{-}=40$ and $E_{+}=80$. From top to bottom, the average time is $5\times10^{5}, 10^{6}, 10^{8}, 10^{9}$, $2\times10^{9}$ and $5\times10^{9}$ time steps, respectively. Inset: Temperature of the left and right log-oscillators (denoted by $T_{-}$ and $T_{+}$) as a function of average time. The parameters for log-oscillators are set by $\tau_{-}=5$, $\tau_{+}=10$ and $\varepsilon=0.0001$.  The coupling parameter between thermostat and system are set by $\gamma_{-}=0.0005$, $\gamma_{+}=0.00005$. The system size $N=52$.}
\end{figure}

In our simulations, the energy of the log-oscillators is initially assigned in the low-energy zone of their own plateau region, namely, $E_{-}=1$ and $E_{+}=20$. As a comparison, we also examine the case that the energy of respective log-oscillators are initially to the edge of the plateau. The initial energy of each FPU oscillator is randomly prepared around an average energy equal to 8. After a relaxation process, the system is coupled with the log-thermostats weakly. The evolution of the temperature profile is shown in Fig.~\ref{Fig6}. One can see that the log-thermostat can drive the system into a nonequilibrium state only for a short time, indicated by a linear temperature gradient. As the average time increases, the log-oscillators cannot maintain the temperature gradient. Our numerical results shows that the FPU system, except the boundary oscillators, reach an energy equilibration after a relaxation time, i.e., heat flow cannot be promoted inside the system. Note that the kinetic temperature of the thermostats is kept at $T_{\pm}=\tau_{\pm}$ for the initial setup that $E_{-}=1$ and $E_{+}=20$, as is expected.  However, if the energy is set near the plateau region ($E_{-}=40$ and $E_{+}=80$), the log-oscillators can easily gain energy from the system and jump out of the plateau region. In this case,  $T_{\pm}$ are quite different with $\tau_{\pm}$. Thus $\tau_{\pm}$ cannot be regarded as measures of the thermostat temperature. Inspired by the Nos\'{e}-Hoover chain thermostats~\cite{Hoover2012book}, we also applied thermostats of a chain of log-oscillators at different temperatures as to see if nonequilibrium steady states can be established. However, this again gives negative results even though the "chaoticity" of the thermostats is increased in this way.  Our results indicate that log-thermostats fail to maintain a nonequilibrium steady state (stationary temperature gradient) even though the energy of log-oscillators are assigned in the plateau region.

In summary, a particular property of the log-oscillator is its infinite heat capacity in terms of the virial theorem, which make it a candidate to serve as a Hamiltonian thermostat. With this study, we demonstrated that kinetic temperature of the modified log-oscillator has a linear dependence on system energy if energy is larger than a critical energy, which indicates a possible violation of the virial theorem. The critical energy increases with average time in a slow way, with which its infinite-time behavior cannot be concluded by numerics. Therefore, it would be safely to argue that virial theorem is practically violated in finite-time numerical simulations. As a counterpart of the virial theorem, GET was further investigated for the thermalized log-oscillator with respect to the canonical distribution. Based on the canonical average, we found that there exists a critical temperature, above which GET is violated, namely, $P_{\mu}$ approaches to an upper bound instead of increasing linearly with the increasing of temperature. Our numerical simulations are consistent with the analysis. Finally, we show that log-thermostats fail to maintain a nonequilibrium steady state although we assign the energy of log-oscillators inside the plateau region. Note that it has been shown that equipartition is not satisfied for a Brownian particle under both harmonic confinement in positional space and logrithmic confinement in momentum space~\cite{Barkai2015PRLb,Barkai2016PRE}. The violation comes from the correction of the the Boltzmann-Gibbs density distribution for such a system. For $T>\tau$, the distribution is unnormalizable~\cite{Lutz2004PRL}. In this sense, one would not expect GET to hold in the first place due to the absence of a normalizable distribution. However, although their model is slightly different with ours, it is still interesting to see that our analysis based on the canonical ensemble average also shows a critical behavior at temperature $T=\tau$, which is consistent with the analysis~\cite{Barkai2016PRE}. Further studies of the deviation from Boltzmann-Gibbs statistics in different situations are thus interesting.
\section*{Methods}
\subsection*{Calculation of Eq.~\eqref{Pe}}
For $T<\tau$, Eq.~\eqref{Pe} gives $P_{\mu}=k_B\tau(1-\Delta)$, where
\begin{equation}\label{Delta}
\begin{split}
\Delta &= \frac{\varepsilon\int_{-\infty}^{+\infty} dx (\varepsilon+x^{2})^{- \frac{\beta \tau}{2} - 1}}{\int_{-\infty}^{+\infty} dx (\varepsilon+x^{2})^{- \frac{\beta \tau}{2}}} \\
&= \frac{ \Gamma (\frac{\beta \tau}{2}) \Gamma (\frac{\beta \tau}{2}+\frac{1}{2})}{\Gamma (\frac{\beta \tau}{2}+1) \Gamma( \frac{\beta \tau}{2}-\frac{1}{2})} \\
&= 1 - \frac{T}{\tau}.
\end{split}
\end{equation}
The last equality in Eq.~\eqref{Delta} is obtained by using the property of Gamma function, i.e., $\Gamma(y+1)=y\Gamma(y)$ for y equal to any positive real number. Therefore, GET with respect to coordinate $x$, i.e. $P_{\mu}=k_{B} T$ is valid for $T<\tau$. For $T\geq\tau$, the participation function is non-normalizable and an analytical form for $P_{\mu}$ is so far unavailable. However, $P_{\mu}$ has an upper bound $k_B\tau$ by noting that $0<\Delta<1$, which indicates that GET is invalid with respect to $x$ provided $T \geq \tau$.


\section*{Acknowledgements}
We acknowledge Jiao Wang and Yong Zhang for helpful discussions and Xiamen Supercomputer Center for
use of its computing facilities. D.H. acknowledges financial support from  NSFC of China (No. 11675133), NSF of Fujian Province (No. 2016J01036), and President Grant of Xiamen University (No. 20720160127). H.Z. acknowledges financial support from NSFC of China (No. 11335006).
\end{document}